\documentclass[
aps,
prx,
twocolumn,
superscriptaddress,
amsfonts,
amsmath
]{revtex4} 

\usepackage{graphicx}
\usepackage[usenames]{color}

\usepackage{braket}

\begin{document}

\title{Finite temperature properties of \\
alternating ferro-antiferromagnetic Heisenberg chains}

\author{Niklas Casper}
\email{n.casper@tu-bs.de}
\affiliation{Institute for Theoretical Physics, Technical University 
Braunschweig, D-38106 Braunschweig, Germany}

\author{Wolfram Brenig}
\email{w.brenig@tu-bs.de}
\affiliation{Institute for Theoretical Physics, Technical University 
Braunschweig, D-38106 Braunschweig, Germany}
\date{\today}


\begin{abstract}

We present a study of the spin-$1/2$ Heisenberg chain with 
alternating ferro- and antiferromagnetic exchange, focusing on the role of the
exchange couplings to cover both, dimer and Haldane limit.
Employing a complementary combination of perturbation theory and 
quantum Monte Carlo simulation, we report results for the 
magnetic susceptibility as well as the dynamic structure factor over a 
wide range of coupling constants and for different temperatures to extract the
spin gap. For a small ferromagnetic coupling, we find good agreement between 
perturbation theory and quantum Monte Carlo. For arbitrary exchange couplings 
we show that the dynamic structure factor, obtained from quantum Monte Carlo, 
scales between triplons and a Haldane chain spectrum.  Finally, we
contrast our findings for the spin gap versus the exchange couplings against 
existing literature.

\end{abstract} 

\maketitle


\section{Introduction}\label{sec:introduction}

Homogeneous nearest-neighbor (NN) spin-$S$ antiferromagnetic chains (AFC) exist
in two variants. Namely gapless for $S=(2\mathbb{N}+1)/2$, versus gapped for
$S=\mathbb{N}$, as conjectured by Haldane \cite{Haldane1983a}. Apart from a
non-degenerate singlet ground for both variants and apart from an expected
difference in their correlation functions, i.e. algebraic versus exponential,
induced by the presence or absence of the spin gap, $S{=}1$ Haldane-chains are
fundamentally different from $S{=}1/2$ AFCs, since the former exhibit a finite
string order parameter (SOP) \cite{Nijs1989, Tasaki1991}. This translates into a
symmetry protected topological ground state of the Haldane chain, and the
existence of edge-modes in open boundary systems, which arise naturally in the
AKLT representation of the ground state \cite{AKLT1987}.

Early on {\it inhomogeneous} $S{=}1/2$ chains have been investigated for their
potential similarities to Haldane chains. In particular bond-alternating
$S{=}1/2$ antiferro-antiferro- and ferro-antiferromagnetic chains (AAC and FAC)
have been under intense scrutiny. First exact diagonalization (ED) studies of
the specific heat and the magnetic susceptibility of the FAC have performed in
Ref.~\cite{Borras1994}.  Extensive analysis of the quantum phase diagram of the
AAC and FAC by density matrix renormalization group (DMRG) has shown almost all
of the parameter space to represent Haldane chain behavior \cite{Watanabe1999},
consistent with bosonization and ED calculations of the SOP \cite{Hida1992}
and findings regarding universality at the magnetic field induced transition
into a Luttinger liquid in the FAC \cite{Sakai1995}.  Excitation spectra and 
dynamic structure factor calculations for the FAC have been performed at zero
temperature, $T{=}0$, by ED on finite systems up to 26 sites \cite{Watanabe1999,
Kokado1999, Paul2017} and by bond operator mean field theory \cite{Paul2017,
Wu1999}. Finite temperature studies of the specific heat, susceptibility and
magnetization of the FAC have also been performed using quantum Monte-Carlo
(QMC) \cite{Aplesnin1999,Yamaguchi2015} including analysis of the spin gap with
\cite{Yamaguchi2015} and without \cite{Aplesnin1999} inter-chain
coupling.

Significant efforts have been made to synthesize spin-$1{/}2$ FACs. In early
materials attempts \cite{Hagiwara1997,Hosokoshi1999}, quasi one-dimensional (1D)
behavior was masked by interchain exchange, inducing magnetic long-range order
(LRO) at rather high temperatures. Subsequent analysis however of the compounds
CuNb$_2$O$_6$ \cite{Kodama1998,Kodama1999}, DMACuCl$_3$
\cite{Ajiro2003,Stone2007,Inagaki2014}, Na$_3$Cu$_2$SbO$_6$
\cite{Miura2006,Miura2008,Kuo2012}, the zinc-verdazyl complex 
C$_{29}$H$_{18}$F$_{12}$N$_5$O$_4$Zn \cite{Yamaguchi2015}, 
and BaCu$_2$V$_2$O$_8$ \cite{Klyushina2018} by thermodynamics, 
including magnetization and magnetic susceptibility measurements, as well as 
nuclear magnetic resonance and inelastic neutron scattering provided clear 
evidence of FAC and possibly also Haldane chain behavior.

Despite these extensive efforts, many open questions still remain to be
investigated for the FAC. From a theoretical point of view this applies in
particular to the {\it finite temperature} spin-dynamics. Therefore, in this
work, we will advance QMC calculations of the momentum resolved dynamic
structure factor $S(k,\omega)$ at $T\neq 0$. Moreover, its features, and in
particular the spin gap will be contrasted against calculations using
perturbation theory and static QMC.  The paper is organized as follows. In
Sec.~\ref{sec:model} we describe the model. Sec.~\ref{sec:method} comprises a
short summary of the quantum Monte Carlo method we use. Following this, in
Sec.~\ref{sec:results}, we detail our results, including the thermodynamic
susceptibility, the dynamic structure factor, and the evolution of the spin
gap. We conclude and summarize our findings in Sec.~\ref{sec:summary}.


\section{Model}\label{sec:model}

The Hamiltonian of the spin-$1{/}2$ FAC reads
\begin{align}
H=  \sum_{i{=}0}^{N{/}2} \left(J_{\text{F}} \, {\bf S}_{2i}\cdot {\bf S}_{2i+1} 
+ 	J_{\text{AF}} \, {\bf S}_{2i+1}\cdot {\bf S}_{2i+2} \right)
  -h  \sum_{i=0}^{N{-}1} S_{i}^z \, .
\end{align}
${\bf S}_i{=}(S_i^x,S_i^y,S_i^z)$ are $S{=}1{/}2$ operators at sites $i$ of a
chain with $N{/}2$ unit cells with periodic boundary conditions
(PBC). The ferro- and antiferromagnetic exchange couplings are labeled by
$J_{\text{F}}{<}0$ and $J_{\text{AF}}{>}0$, with a dimensionless parameter 
$j=|J_{\text{F}}/J_{\text{AF}}|$
used hereafter. $h$ refers to an external magnetic field. 
With $N=4 \mathbb{N}$ spins, the system is frustration-free.

In the limiting case of $j=0$, the chain consists of $N/2$ decoupled
antiferromagnetic dimers with a singly degenerate singlet-product ground state
and an energy gap of $\Delta = J_{\text{AF}}$ to a $3 N/2$-fold degenerate set of
first excited triplets states. For $j\neq 0$, but still $j\ll 1$ the latter set
splits into a gas of dispersive and interacting triplon excitations. In the
opposite limit, i.e. $1/j = 0$ the system comprises $N/2$ decoupled
ferromagnetic dimers in triplet states with a $3^{N/2}$-fold ground state
degeneracy and an energy gap of $\Delta = J_{\text{F}}$ to an $N/2$-fold 
degenerate set of first excited singlet states. For $1/j\neq 0$, but still 
$1/j\ll 1$ the ground state degeneracy is lifted and the triplets are coupled 
into an effective low-energy antiferromagnetic spin-$1$ chain, i.e. 
the Haldane chain \cite{Haldane1983a}.

Starting with early ED-work \cite{Botet1983}, more recent analysis using QMC
\cite{Nightingale1986} and DMRG \cite{White1993} has converged to a spin gap of
$\Delta_H/J\simeq 0.41050(2)$ for the Haldane chain, where $J$ is the exchange
coupling constant. It is tempting to identify the latter $J$ with 
$J_{\text{AF}}$ of the FAC in the asymptotic situation $1/j\rightarrow 0$, 
but $J_{\text{AF}}\neq 0$, and therefore to also expect a gap of 
$\Delta/J_{\text{AF}}\simeq 0.41$ for the FAC~\cite{Aplesnin1999}. However, 
since bond-correlation functions on the AF bonds
in the FAC, i.e. $\langle {\bf S}_{2i-1}\cdot {\bf S}_{2i}\rangle$ are reduced
by a factor of 4 as compared to the bond-correlation functions of a fictitious
Haldane chain \cite{Hung2005} with spin $L=1$, i.e.  $\langle {\bf L}_{i}\cdot
{\bf L}_{i+1}\rangle = 4 \langle {\bf S}_{2i}\cdot {\bf S}_{2i+1}\rangle$, one
instead should expect a rescaling of the spin gap by a factor of $1/4$.

\begin{figure}
\centering
\includegraphics[width=.7\columnwidth]{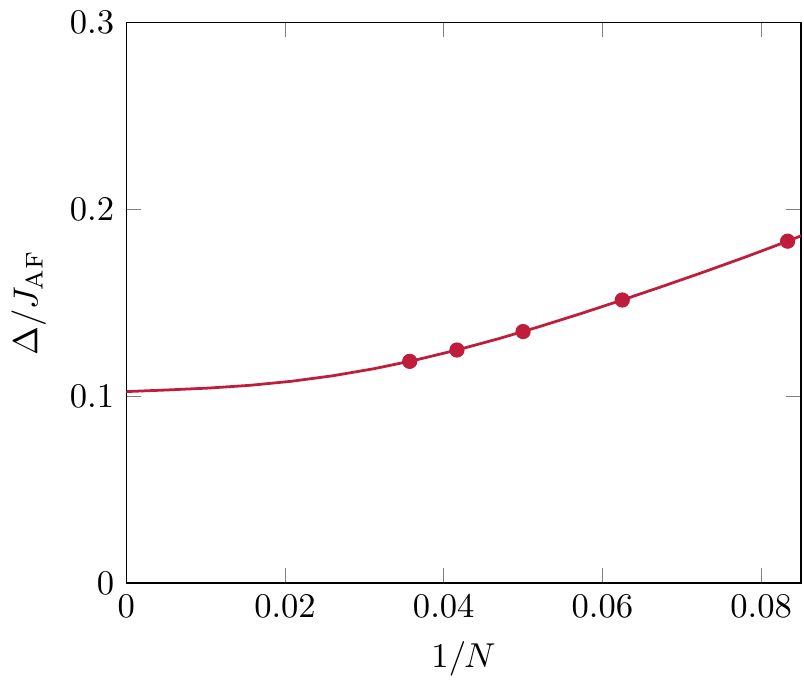}
\caption{Finite size scaling of spin gap (solid dots) from Lanczos
\cite{Bauer2011} for $N=12\dots 28$. Solid line is a non-linear fit.}
\label{fig:Lanczos}
\end{figure}

We emphasize the previous point in two ways. First, in Table~\ref{table:factor}
we compare the spectra of two toy models, namely an $L=1$ AF dimer and a FAC
chain of length $N=4$, both with open boundary conditions. By defining the gap 
as the difference of energies between the lowest triplet and singlet states, 
an exact reduction of the gap by $1/4$ is obvious. Second, 
in Fig.~\ref{fig:Lanczos} we show a polynomial fits in $1{/}N$ to small 
system diagonalization using Lanczos from the ALPS project \cite{Bauer2011}. 
Already for these very short chains ($N=12\dots 28$) an  extrapolated value of 
${\Delta/J_{\text{AF}}\approx 0.1025 = 0.41/4}$ is manifest.

\begin{table}
\begin{tabular}{lrc}
\hline\hline
& \multicolumn{1}{c}{$S=1$} & $S=1/2$\\\hline
singlet $E_0$ & $-2J_{\text{AF}}$ & $-1/4 (1-2j+2\sqrt{1+2j+4j^2})$\\
triplet $E_1$ & $-J_{\text{AF}}$ & $-1/4 (1+2\sqrt{1+j^2})$\\
gap $\Delta=E_1-E_0$ & $J_{\text{AF}}$ & 
	$\lim\limits_{j\to\infty}\Delta=J_{\text{AF}}/4$\\\hline\hline
\end{tabular}
\caption{Spectra of $S{=}1$ dimer and $N{=}4$, $S{=}1{/}2$ FAC}
\label{table:factor}
\end{table}


\section{Quantum Monte Carlo method}\label{sec:method}

The numerical results obtained in this work are based on QMC 
calculations using the stochastic series expansion (SSE), as pioneered 
in Refs. \cite{Sandvik1992, Sandvik1999, Syljuasen2002}. 
This method is based on an importance sampling of the
high temperature series expansion of the partition function
\begin{equation}\label{eq:partition}
Z=\sum_{\alpha}\sum_{S_{M}}\frac{(-\beta)^{n}(M-n)!}{M!}
\Braket{\alpha|\prod_{p=1}^{M}H_{a_{p},b_{p}}|\alpha} \, ,
\end{equation}
where $\beta{=}1/T$ is the inverse temperature, 
${H_{1,b} = C-S_{b1}^{z} S_{b2}^{z}}$ and 
${H_{2,b} = (S_{b1}^{+} S_{b2}^{-} + S_{b1}^{-} S_{b2}^{+})/2}$
are the spin diagonal and off-diagonal bond operators, 
and $M$ the truncation order. 
$C$ must be chosen such that all diagonal weights are nonnegative. 
${\Ket{\alpha} = \Ket{S_{1}^{z}, \ldots,S_{N}^{z}} }$ 
refers to the $S^{z}$ basis and
${S_{M}=[a_{1},b_{1}][a_{2},b_{2}]\ldots[a_{M},b_{M}]}$ 
is an index for the so-called operator string $\prod_{p=1}^{M}H_{a_{p},b_{p}}$. 
This string is Metropolis sampled, using two types of updates, 
i.e. diagonal updates which change the number of diagonal operators 
$H_{1,b_{p}}$ in the operator string and
loop updates which change the type of operators $H_{1,b_{p}}\leftrightarrow
H_{2,b_{p}}$. For bipartite lattices the loop update comprises an even number of
off-diagonal operators $H_{2,b_{p}}$, ensuring positivity of the transition
probabilities. The order $M$ of the expansion is truncated depending on $T$,
such as to have no impact on precision.

The dynamic structure factor can be obtained from QMC in real space by
a conversion of the discrete expansions slices to continuous imaginary
time via a binomial distribution \cite{Sandvik1992}

\begin{eqnarray}
\Braket{S_i(\tau)S_j(0)}=
\left\langle \sum_{m=0}^{M} {M \choose m} 
\left(\frac{\tau}{\beta}\right)^m\nonumber 
\left(1-\frac{\tau}{\beta}\right)^{M-m}\nonumber \right. \\
\left. \frac{1}{M} \sum_{p=0}^{M-1} S_{i}^{+}(m+p)S_{j}^{-}(p)
\right\rangle_{W} \, ,
\label{a2}
\end{eqnarray}
where $i$, $j$ refer to sites, and $\tau$ to the imaginary time. $m+p$, $p$ on
the right hand side label positions within the operator string, and 
$\langle \dots \rangle_{W}$ denotes the Metropolis weight of an operator string 
of length $M$ generated by the SSE \cite{Sandvik1999, Syljuasen2002}. 
From Eq.~(\ref{a2}) one can proceed to momentum space by Fourier transformation
\begin{equation}\label{eqn:structureFactor}
S(k, \tau) =\sum_{i}e^{\mathrm{i} k r_i} \Braket{S_i(\tau)S_0(0)}/N \, .
\end{equation}
Since the model comprises two sites per unit cell we evaluate an even(odd)
structure factor $S_{even(odd)}(k,\omega)$ by summing over $i{=}2 l (i{=}2l+1)$
in Eq.~(\ref{eqn:structureFactor}) including the on-site/bond correlators. 
Finally, the dynamic structure factor in frequency and momentum space is 
obtained from analytic continuation, which is equivalent to an inversion for 
$S(k ,\omega)$ of

\begin{equation}
S(k, \tau) =\int_{0}^{\infty} 
d\omega \, S(k ,\omega)K(\omega,\tau) \, ,
\end{equation}

with a kernel $K(\omega,\tau)=(e^{-\tau\omega}+e^{-(\beta-\tau)\omega})/\pi$.

The preceding inversion is an ill-posed problem, for which maximum entropy
methods (MEM) have proven to be well suited. We have used Bryan's MEM algorithm
\cite{Skilling1984, Jarrell1996}. This method minimizes the functional
$Q=\chi^{2}/2-\alpha\sigma$, with $\chi$ being the covariance of the QMC data
with respect to the MEM trial spectrum $S(k, \omega)$. Overfitting is prevented 
by an entropy term 
$\sigma=\sum_{\omega} S(k, \omega) \ln[S(k, \omega) /m(\omega)]$.  
We have used a flat default model $m(\omega)$, which is iteratively adjusted to 
match the zeroth moment of the trial spectrum. The optimal spectrum follows 
from the average of $S(k, \omega)$, weighted by a probability distribution 
$P[\alpha| S(k, \omega)]$ (see \cite{Skilling1984}).


\section{Results}\label{sec:results}

In this section, we will detail our results for the finite temperature
susceptibility and the dynamic structure factor. For the former we will contrast
perturbation theory (PT) with QMC. One of the quantities of prime interest to be
extracted from this is the spin gap, which we will discuss also.

\begin{figure}
\centering
\includegraphics[width=\linewidth]{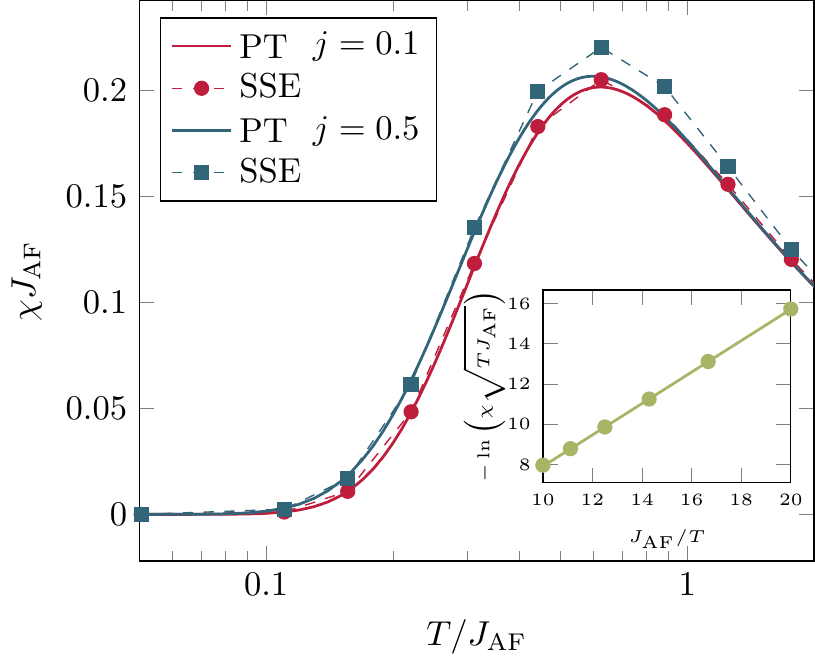}
\caption{
Susceptibility versus $T$ for $j$=0.1 and 0.5 obtained from PT (solid) and SSE
(dashed with markers). For SSE the error is less than the marker size. Inset:
Typical fits of Eq.~(\ref{eq:fit}) (solid) to SSE (markers) in the low-$T$ range
for $j=2$, where the gap $\Delta$ can be extracted from the slope.
}
\label{fig:susceptibility}
\end{figure}

\subsection{Magnetic susceptibility}

\noindent
{\it \underline{Perturbation theory} (PT):} For $j\ll 1$ we first consider the
coupling between the AF dimers perturbatively. Following Ref.~\cite{James2008},
we expand to $O(j^1)$. This turns the problem into a {\em tight-binding} model
for a gas of {\it non-interacting} triplets with nearest-neighbor hopping along
the chain (triplons) which acquire a dispersion of
\begin{equation}
\epsilon_{k} = J_{\text{AF}} - \frac{J_{\text{F}}}{2} \cos (k a) 
	= \Delta - \frac{J_{\text{F}}}{2} \left(1 + \cos (k a) \right)\, ,
\label{ptdisp}
\end{equation}
where the length of the unit cell $a \equiv 2$ is set out
hereafter. $\Delta = \epsilon_{\pi/a} = J_{\text{AF}} + \frac{J_{\text{F}}}{2}$ 
is the gap. The free energy of such triplon gases is given by \cite{Troyer1994}
\begin{align}
 f&=-\frac{1}{2\beta} \ln\big\{ 1+ [1+2\cosh(\beta h)]
\sum_k  e^{-\beta \epsilon_k }\big\}\, .
\end{align}
The momentum summation $z(\beta)=\sum_k  e^{-\beta \epsilon_k }$ can be
evaluated as
\begin{equation}
 z(\beta) =\frac{1}{2\pi}\int_{-\pi}^{\pi} dk  \, e^{-\beta \epsilon_k }
 = e^{\beta J_{\text{AF}}} I_0\left(\frac{\beta J_{\text{F}}}{2}\right) \, ,
\label{zbeta}
\end{equation}
where $I_0(x) = \sum_n ( x^2/4 )^n/(n!)^2$ is the modified Bessel function of 
the first kind \cite{Abramowitz1970}. From this the susceptibility can be 
obtained analytically by
\begin{equation}\label{eqn:susceptibility}
\chi(T) =
-\frac{\partial f}{\partial h^2}|_{h=0} =
\beta  \frac{z(\beta)}{1+3z(\beta)} \, .
\end{equation}

\begin{figure*}[tb]
\centering
\includegraphics[width=0.9\textwidth]{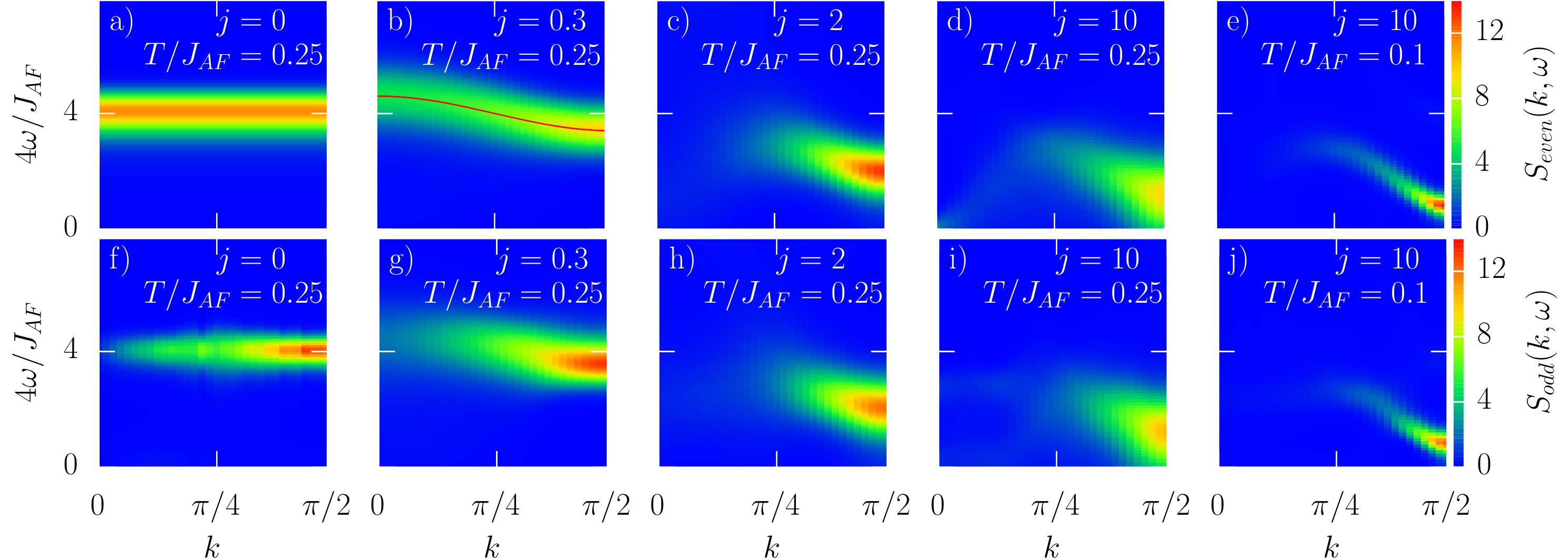}
\caption{Contour maps of the dynamic structure factors $S_{even / odd} (k,
\omega)$ for various $j$=0, 0.3, 2, and 10, as well as for temperatures
$T/J_{\text{AF}}$=0.25 and 0.1 on a FAC with $N=128$ sites. Solid line in 
panel b) refers to PT, Eq.~(\ref{ptdisp}). Color coding of intensity 
(bar on rightmost panels) identical in all plots.}
\label{fig:dynamic}
\end{figure*}

\noindent
{\it \underline{SSE quantum Monte Carlo}:} 
The susceptibility can be calculated within the SSE method by measuring the sum
over the spin configurations
\begin{equation}
\chi = \frac{1}{T}
\left(\langle (\sum_i S_i^z)^2 \rangle - \langle\sum_i S_i^z\rangle^2\right) \,.
\label{eq:chi}
\end{equation}
For low temperatures the energy gap between the ground state and the first
excited state can be extracted according to \cite{Damle1998}
\begin{equation}
\chi (T) \propto \sqrt{\beta} \, e^{- \beta \Delta} \, ,
\label{eq:fit}
\end{equation}
which stems from the asymptotic expansion of the modified Bessel function
for large arguments \cite{Abramowitz1970} in Eq.~(\ref{eqn:susceptibility}).

Fig.~\ref{fig:susceptibility} shows $\chi(T)$ for a low and an intermediate
temperature range and for small, as well as for intermediate $j$=0.1 and 0.5,
respectively. As is obvious PT and SSE agree very well for 
$T/J_{\text{AF}} \ll 1$ and for both values of $j$. Interestingly, 
this agreement holds up to $j$=0.5 even at intermediate $T/J_{\text{AF}}$, 
where only a $\sim 10\%$ difference can be observed at the correlation maximum. 
Because of the excellent agreement at low $T/J_{\text{AF}}$,
fitting the SSE with Eq.~(\ref{eq:fit}) is well justified. We note in passing
that for $T/J_{\text{AF}}\gg 1$, PT and SSE both approach Curie behavior, with
$\chi(T)=C_c/T$, and a Curie constant $C_c=1/4$ per spin. 

\subsection{Dynamic structure factor}

Now we turn to the finite-temperature dynamic structure factor.
Fig.~\ref{fig:dynamic} displays contour maps of $S_{even/odd}
(k,\omega)$ over a wide range of $j$-values and for two temperatures. The figure
conveys four main messages.

First, there is an obvious evolution of dispersive behavior starting with a 
$k$-independent gap at $j$=0. The latter simply reflects the spectrum of
the exact AF-dimer product-state, which comprises delta-functions at the
singlet-triplet gap $\omega=J_{\text{AF}}$. Increasing $j$, one remains 
within the range of validity of PT, where we expect $S_{even/odd} (k,\omega)$ 
to encode a cosine-band of dispersing triplets. As is shown in 
Fig.~\ref{fig:dynamic}b) the center of gravity of the spectrum fits perfectly to 
Eq.~(\ref{ptdisp}) indeed. Finally, as $j$ is increased further, the system 
crosses over into a correlated state displaying a small but {\it finite} gap 
at which the Brillouin zone (BZ) boundary intensity is maximal. We find, 
that increasing $j$ beyond the value of $j^\star{\sim} O(10)$, has little effect
on the size of this gap. We conclude, that $j>j^\star$ is the regime of 
the effective Haldane chain.

Second, Fig.~\ref{fig:dynamic} displays a gradual redistribution of spectral
weight for $S_{even}$ with increasing $j$, 
starting from a situation with significant intensity
extending over all of the BZ at $j\ll 1$, to a prominent modulation of the 
spectral weight located at the zone boundary, occurring for $j\gg 1$.
This is a direct manifestation of the increase of the AF spin-correlation length
$\xi$ versus $j$ as the Haldane limit is approached. In fact for $j=0$, AF
correlations extend only over a single dimer, while in the Haldane limit
$\xi\simeq 6$, referring to $S$=1-sites \cite{White1993}, i.e. FM dimers in our
case. This increase of $\xi$ translates into a static structure factor $S_k \sim
\int^\infty_0 d\omega S(k,\omega)$, which evolves from a momentum independent function
at $j{=}0$ to one which displays an increase as $k\rightarrow\pi/2$ for
$j\rightarrow\infty$. This causes the intensity modulation, observable
in $S_{even}(k,\omega)$.
For $S_{odd}$, already at $j\ll 1$ the spectral weight is located at the 
zone boundary because of an additional trivial modulation by 
on-dimer correlations as described in Eq.~\eqref{eqn:structureFactor}.

Third, there is a clear thermal broadening of the spectra as $T$
increases. This can be seen in Fig.~\ref{fig:dynamic}d),~e) and i),~j) at
$j$=10. In both cases the $T$=0.1 spectra are rather sharp. They display a clear
gap even for $j$=10 and very much resemble that of the Haldane chain at $T=0$
\cite{Takahashi1989,Golinelli1993,White1993,Grossjohann2010,Rahnavard2015}.
Increasing $T/J_{\text{AF}}$ to 0.25, keeping $j$ fixed, 
the spectra a broadened and the gap is practically closed. 
Again this is similar to finite temperature analysis of
the pure Haldane chain \cite{Grossjohann2010,Rahnavard2015,Richter2019}.

Fourth. there is only a small difference between $S_{even/odd} (k,\omega)$ in 
a intensity modulation along $k$. 
As described in Eq.~(\ref{eqn:structureFactor}), 
the on-site/bond correlators are included and are leading to this difference 
clearly seen for $j=0$.

Out of scale of Fig.~\ref{fig:dynamic} is another band at the 
ferromagnetic dimer $\omega = j$ with low intensity. 
Only for $j\geq 2$ this band is well seperated 
from the dispersive band. In MEM therefore, the whole spectrum and thus a larger 
$\omega$-regime must be considered.

\subsection{Gap evolution}

\begin{figure}[tb]
\centering
\includegraphics[width=\linewidth]{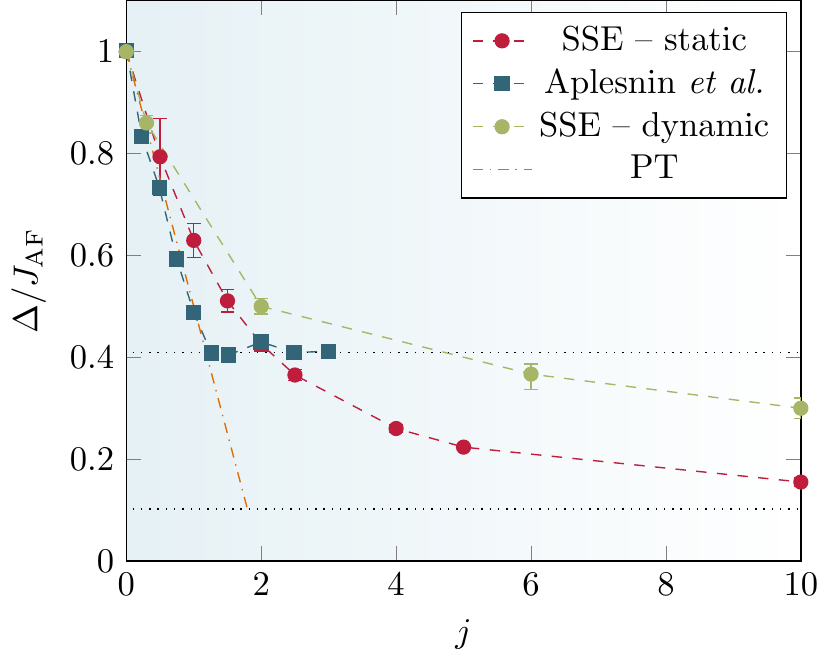}
\caption{Spin gap versus $j$, obtained from various
methods: PT, thermodynamic and dynamic QMC. Dotted horizontal lines:
conventional Haldane gap $\approx0.41$ and one forth thereof. For dynamic QMC 
the error is the full width at half maximum.}
\label{fig:gap}
\end{figure}

In this subsection we summarize our findings for the spin gap versus $j$.  To
this end, in Fig.~\ref{fig:gap}, we have collected $\Delta/J_{\text{AF}}$ 
as obtained from our PT, thermodynamic SSE, and dynamic SSE calculations.  
For dynamic SSE, the gap is defined by the energy $\omega_{max}/J_{\text{AF}}$ 
at the maximum of  $S(k, \omega)$ at $k=\pi/2$. In addition, the figure also 
contains QMC data from Aplesnin and Petrakovskii \cite{Aplesnin1999}.

First, it is obvious, that for $j\lesssim 1$ all methods result in a gap of
comparable magnitude, which decreases rapidly with $j$. Second, for $j\gtrsim 1$
PT underestimates the gap size. Third, thermodynamic and dynamic SSE both show
the same trend of the gap, i.e. to converge to a value which is clearly less
than the single spin-1 Haldane chain. For $\Delta$ as from thermodynamic SSE and
Eq.~(\ref{eq:fit}) is tempting to speculate, that indeed
$\Delta(j\rightarrow\infty) = \Delta_H/4$ as conjectured in Section
\ref{sec:model}. The gap from dynamic SSE shows a similar trend, however the
convergence to the anticipated value of $\Delta_H/4$ is significantly
slower. Since $\chi(T)$ rather refers to a measure of an integrated density of
states, than a peak position, a {\it quantitative} difference in $\Delta$ as
obtained from thermodynamic and dynamic QMC is not surprising. Turning to
Ref.~\cite{Aplesnin1999}, where the gap has also been extracted from fits to
$\chi(T)$, obtained by a QMC method, we note that agreement with our results is
visible for $\Delta \geq \Delta_H$, i.e. for $j\lesssim 1$. However beyond that,
$\Delta$ from Ref.~\cite{Aplesnin1999} is pinned to $\Delta_H$. This is a
variance not only with our results, but also with Ref.~\cite{Hung2005}. 
The cause of this remains unclear at present.

\section{Conclusion}\label{sec:summary}

To summarize, we have studied static and dynamic properties of alternating
ferro-antiferromagnetic spin-1/2 chains at finite temperature. We observe a
smooth crossover from weakly coupled dimer to Haldane-chain behavior versus $j$
with {\it no} intervening quantum critical point. Both, thermodynamic and
dynamic QMC suggest that lowest-order PT for a dispersive triplon gas describes
the physics well up to $j \sim 1$. At $j\gtrsim 10$ effective Haldane physics
emerges, with an asymptotic spin gap reduced by a factor of 4 as compared to the
conventional Haldane gap. Because of the latter reduction, the impact of finite
temperature is significant already a $T/J_{\text{AF}} \gtrsim 0.25$, 
where we find the gap to be filled in completely. 

While we have focused on periodic boundary conditions, open chains can equally
well be studied by SSE QMC. This would allow to analyze the fate of topological
edge states versus $j$ and temperature. In view of the existing materials,
C$_{29}$H$_{18}$F$_{12}$N$_5$O$_4$Zn \cite{Yamaguchi2015} and BaCu$_2$V$_2$O$_8$
\cite{Klyushina2018}, more realistic dynamic QMC calculations, including inter
chain exchange should be an additional direction of future research.


\acknowledgments
This work has partially been supported by the State of Lower Saxony through
QUANOMET (project NP-2) and by the DFG via SFB 1143 (Project
A02). W.~B. acknowledges kind hospitality of the PSM, Dresden.



\begin{thebibliography}{99}


\bibitem{Haldane1983a}
F. D. M. Haldane, Phys. Lett. A {\bf 93}, 464 (1983).

\bibitem{Nijs1989}
M. den Nijs and K. Rommelse, Phys. Rev. B {\bf 40}, 4709 (1989).

\bibitem{Tasaki1991}
H. Tasaki, Phys. Rev. Lett. {\bf 66}, 798 (1991).

\bibitem{AKLT1987}
I. Affleck, T. Kennedy, E. H. Lieb, H. Tasaki,
Phys. Rev. Lett. {\bf 59}, 799  (1987).

\bibitem{Borras1994}
J. J. Borras-Almenar, E. Coronado, J. Curely, R. Georges, and
J. C. Gianduzzo, Inorganic Chemistry {\bf 33}, 5171 (1994).

\bibitem{Watanabe1999}
S. Watanabe and H. Yokoyama,
J. Phys. Soc. Jpn. {\bf 68}, 2073 (1999).

\bibitem{Hida1992}
K. Hida,
Phys. Rev. B {\bf 45}, 2207 (1992).

\bibitem{Sakai1995}
T. Sakai, J. Phys. Soc. Jpn. {\bf 64}, 251 (1995).

\bibitem{Kokado1999}
S. Kokado and N. Suzuki,
J. Phys. Soc. Jpn. {\bf 68}, 3091 (1999).

\bibitem{Paul2017}
S. Paul and A. K. Ghosh,
Condensed Matter Physics {\bf 20}, 23701 (2017).

\bibitem{Wu1999}
Y.-Z. Wu and Z.-Y. Li,
Physica Status Solidi (B) {\bf 213}, 27 (1999).

\bibitem{Aplesnin1999}
S. S. Aplesnin and G. A. Petrakovskii, Phys. Solid State {\bf 41}, 1511 (1999).

\bibitem{Yamaguchi2015}
H. Yamaguchi, Y. Shinpuku, T. Shimokawa, K. Iwase, T. Ono, Y. Kono,
S. Kittaka, T. Sakakibara, and Y. Hosokoshi, Phys. Rev. B {\bf 91}, 
085117 (2015).


\bibitem{Hagiwara1997}
M. Hagiwara, Y. Narumi, K. Kindo, T. C. Kobayashi, H. Yamakage, K. Amaya,
and G. Schumauch, J. Phys. Soc. Jpn. {\bf 66}, 1792 (1997).

\bibitem{Hosokoshi1999}
Y. Hosokoshi, Y. Nakazawa, K. Inoue, K. Takizawa, H. Nakano, M. Takahashi,
and T. Goto, Phys. Rev. B {\bf 60}, 12924 (1999).

\bibitem{Kodama1999}
K. Kodama, H. Harashina, H. Sasaki, M. Kato, M. Sato, K.
Kakurai, and M. Nishi, J. Phys. Soc. Jpn. {\bf 68}, 237 (1999).

\bibitem{Inagaki2014}
Y. Inagaki, Y. Sakamoto, H. Morodomi, T. Kawae, Y. Yoshida,
T. Asano, K. Hosoi, H. Kobayashi, H. Kitagawa, Y. Ajiro, and
Y. Furukawa, J. Phys. Soc. Jpn. {\bf 83}, 054716 (2014).

\bibitem{Kodama1998}
K. Kodama, T. Fukamachi, H. Harashina, M. Kanada,
Y. Kobayashi, M. Kasai, H. Sasai, M. Sato, and K. Kakurai,
J. Phys. Soc. Jpn. {\bf 67}, 57 (1998).

\bibitem{Ajiro2003}
Y. Ajiro, K. Takeo, Y. Inagaki, T. Asano, A. Shimogai, M. Mito,
T. Kawae, K. Takeda, T. Sakon, H. Nojiri, and M. Motokawa,
Physica B {\bf 329}-{\bf 333}, 1008 (2003).

\bibitem{Stone2007}
M. B. Stone, W. Tian, M. D. Lumsden, G. E. Granroth, D.
Mandrus, J.-H. Chung, N. Harrison, and S. E. Nagler, Phys.
Rev. Lett. {\bf 99}, 087204 (2007).

\bibitem{Miura2006}
Y. Miura, R. Hirai, Y. Kobayashi, and M. Sato, J. Phys. Soc. Jpn.
{\bf 75}, 084707 (2006).

\bibitem{Miura2008}
Y. Miura, Y. Yasui, T. Moyoshi, M. Sato, and K. Kakurai,
J. Phys. Soc. Jpn. {\bf 77}, 104709 (2008).

\bibitem{Kuo2012}
C. N. Kuo, T. S. Jian, and C. S. Lue,
J. Alloys Compd. {\bf 531}, 1 (2012).

\bibitem{Klyushina2018}
E. S. Klyushina, A. T. M. N. Islam, J. T. Park, E. A. Goremychkin, E. Wheeler,
B. Klemke, and B. Lake, Phys. Rev. B {\bf 98}, 104413 (2018).


\bibitem{Botet1983}
R. Botet and R. Jullien, Phys. Rev. B {\bf 27}, 613 (1983).

\bibitem{Nightingale1986}
M. P. Nightingale and H. W. J. Blöte, Phys. Rev. B {\bf 33}, 659 (1986).

\bibitem{White1993}
S. R. White and D. A. Huse, Phys. Rev. B {\bf 48}, 3844 (1993).

\bibitem{Hung2005}
H.-H. Hung and C.-D. Gong, Phys. Rev. B {\bf 71}, 054413 (2005).

\bibitem{Bauer2011}
B. Bauer, L. D. Carr, H. G. Evertz, {\it et al.}, J. Stat. Mech. P05001 (2011).


\bibitem{Sandvik1992}
A. W. Sandvik, J. Phys. A {\bf 25}, 3667 (1992).

\bibitem{Sandvik1999}
A. W. Sandvik, Phys. Rev. B {\bf 59}, R14157 (1999).

\bibitem{Syljuasen2002}
O. F. Sylju\aa{}sen and A. W. Sandvik, Phys. Rev. E {\bf 66}, 046701 (2002).

\bibitem{Skilling1984}
J. Skilling and R. K. Bryan, Mon. Not. R. Astron. Soc. {\bf 211}, 111 (1984).

\bibitem{Jarrell1996}
M. Jarrell and J. Gubernatis, Phys. Rep. {\bf 269}, 133 (1996).



\bibitem{James2008}
A. J. A. James, F. H. L. Essler, and R. M. Konik, Phys. Rev. B {\bf 78}, 094411 
(2008).

\bibitem{Troyer1994}
M. Troyer, H. Tsunetsugu, and D. W\"urtz, Phys. Rev. B {\bf 50}, 13515 (1994).

\bibitem{Abramowitz1970}
M. Abramowitz, Handbook of mathematical functions: with formulas, graphs, and 
mathematical tables (Dover Publications, New York, 1970).

\bibitem{Damle1998}
K. Damle and S. Sachdev, Phys. Rev. B {\bf 57}, 8307 (1998).


\bibitem{Takahashi1989}
M. Takahashi, Phys. Rev. Lett. {\bf 62}, 2313 (1989).

\bibitem{Golinelli1993}
O. Golinelli, T. Jolicoeur, and R. Lacaze, J. Phys.: Condens. Matter {\bf 5}, 
1399 (1993).

\bibitem{Grossjohann2010}
S. N. Grossjohann, Static and Dynamic Properties of Low Dimensional Quantum Spin
Systems, dissertation, Technische Universit\"at Braunschweig, Cuvillier Verlag
G\"ottingen, 2010.

\bibitem{Rahnavard2015}
Y. Rahnavard and W. Brenig, Phys. Rev. B {\bf 91}, 054405 (2015).

\bibitem{Richter2019}
J. Richter, N. Casper, W. Brenig and R. Steinigeweg, 
Phys. Rev. B {\bf 100}, 144423 (2019).

\end{thebibliography}
\end{document}